\documentclass[aps,prb,twocolumn,amsmath,amssymb,nofootinbib,floatfix,superscriptaddress]{revtex4-1}

\usepackage{amsmath,braket}
\usepackage{amssymb}
\usepackage{amsthm}
\usepackage[dvipdf]{color}
\usepackage{graphicx}
\usepackage{dcolumn}
\usepackage{bm,subfigure} 
\usepackage{xcolor,colortbl}
\usepackage[normalem]{ulem}

\DeclareMathOperator{\arccosh}{arccosh}

\begin{document}

\title{Band theory and boundary modes of high-dimensional representations of infinite hyperbolic lattices}

  \author{Nan Cheng}
  \author{Francesco Serafin}
  \author{James McInerney}
 \affiliation{
 Department of Physics,
  University of Michigan, Ann Arbor, 
 MI 48109-1040, USA
 }
 \author{Zeb Rocklin}
 \affiliation{
 School of Physics, Georgia Institute of Technology, Atlanta, Georgia, 30332, USA
 } 
 \author{Kai Sun}
  \author{Xiaoming Mao}
 \affiliation{
 Department of Physics,
  University of Michigan, Ann Arbor, 
 MI 48109-1040, USA
 }

\begin{abstract} 
Periodic lattices in hyperbolic space are characterized by symmetries beyond Euclidean crystallographic groups, offering a  new platform for classical and quantum waves, demonstrating great potentials for a new class of topological metamaterials. One important feature of hyperbolic lattices is that their translation group is nonabelian, permitting high-dimensional irreducible representations (irreps), in contrast to abelian translation groups in Euclidean lattices. 
Here we introduce a general framework to construct wave eigenstates of high-dimensional irreps of infinite hyperbolic lattices, thereby generalizing Bloch's theorem, and discuss its implications on unusual mode-counting and degeneracy, as well as bulk-edge correspondence in hyperbolic lattices. We apply this method to a mechanical hyperbolic lattice, and characterize its band structure and zero modes of high-dimensional irreps.
\end{abstract}
\maketitle

\noindent{\it Introduction}---Bloch’s theorem has been the foundation of solid state physics. From the concept of energy bands to the blossoming field of topological insulators, everything starts with how wave eigenstates are modulated by spatially periodic potentials in crystals.  The abelian nature of the translation groups in crystals limits their representations to one-dimensional (1D), i.e., the Bloch factor, $e^{ikr}$, greatly simplifying the mathematical description of waves in crystals.

New materials and structures with complex spatial order beyond periodic lattices are being discovered, with a  particularly interesting class being hyperbolic lattices, which have recently evolved from pure mathematical concepts~\cite{coxeter1954regular} to real materials realizable in the lab~\cite{kollar2019hyperbolic,kollar2020line,Boettcher2020,Yu2020,Maciejko2021,RUZZENE2021101491,lenggenhager2021electric,Maciejko2022,Boettcher2022,Urwyler2022hyperbolic,Bienias2022}.  These lattices are perfectly ordered in hyperbolic space, i.e., space with constant negative curvature. 
A simple example of a 2D hyperbolic lattices is the tiling of regular heptagons where three heptagons meet at each vertex (i.e., the $\{7,3\}$ tiling). The interior angle of a heptagon in a flat plane is greater than $2\pi/3$, leading to an obvious frustration.  This frustration is resolved on a hyperbolic plane, where the interior angle is modified by the Gaussian curvature. Interestingly, in contrast to limited choices of regular lattices in Euclidean space, there are infinitely many regular lattices in hyperbolic space, opening up a huge space for unconventional symmetries and physics.

How to describe waves in hyperbolic lattices? In recent studies, a range of intriguing features have been reported, e.g., topological edge states~\cite{Yu2020,Urwyler2022hyperbolic}, higher-genus torus Brillouin zones (BZs)~\cite{kollar2019hyperbolic,Maciejko2021,Maciejko2022,kollar2020line}, and circuit quantum electrodynamics~\cite{Boettcher2020,Bienias2022}, outlining an exciting arena of new theories. 
However, key questions still remain on fundamental principles of constructing wave eigenstates from the symmetries of these hyperbolic lattices.  As mentioned above, the simple form of the Bloch factor comes from the abelian translation group in Euclidean space.  In hyperbolic lattices, in contrast,  translations form an infinite non-abelian group, calling for high-dimensional irreps.  How to construct waves of these high-dimensional irreps, and the fundamental physics of the resulting waves, remain open questions. 
An alternative way to demonstrate the necessity of high-dimensional irreps in hyperbolic lattices is the scaling of the number of wave modes with the system size. A Euclidean lattice of linear size $L$ with $n$ degrees of freedom (DOFs) per unit cell has $n$ bands in reciprocal ($k$) space, and the number of points in the first BZ is $L^D$ (where $D$ is the spatial dimension), making the numbers of DOFs in real space and $k$-space equal.  In a hyperbolic lattice, however, the number of unit cells grows \emph{exponentially} with $L$, leading to $\mathcal{O}(e^L)$ wave modes, which is much greater than the number of points in the first BZ.  As we analyze in this letter, this indicates that a sequence of high-dimensional irreps is needed to define a complete basis of waves on hyperbolic lattices at large $L$.

In this letter, we introduce a generalized Bloch's theorem for high-dimensional irreps of the non-abelian translation groups of infinite hyperbolic lattices, which allows us to construct wave eigenstates for any given high-dimensional irreps on hyperbolic lattices, and we discuss the unusual physics of these waves. 
We find that $d \times n$ bands of bulk waves arise from  $d$-dimensional unitary irreps in hyperbolic lattices (in contrast to $n$ bands in Euclidean lattices).   In addition, spatially localized edge/interface modes must involve high-dimensional irreps. 
We apply this formulation to a hyperbolic mechanical lattice, and reveal a series of unusual features from zero modes in hyperbolic lattices where the bulk is over-constrained, to a modified bulk-edge correspondence for potential topological states in hyperbolic space.

\begin{figure}
    \centering
    \includegraphics[width=0.48 \textwidth]{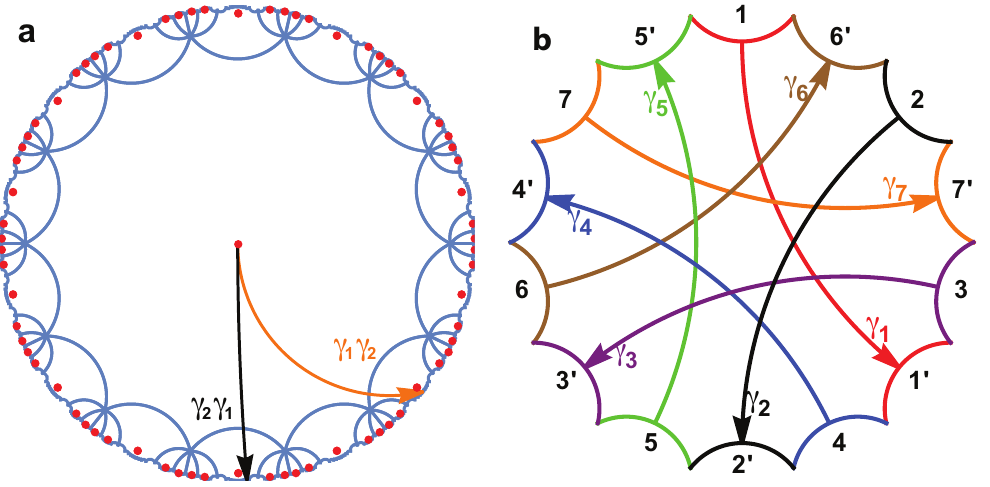}
    \caption{The Bravais lattice and translation group of a 2D hyperbolic periodic tiling $\{14,7\}$. (a) The $\{14,7\}$ tiling (blue geodesics showing the edges) and its dual lattice $\{7,14\}$ (red dots showing the nodes)  on the Poincar\'e disk. The $\{7,14\}$ is a 
    generalized Bravais lattice. 
    The two arrows mark translations  $\gamma_2\gamma_1$ and $\gamma_1\gamma_2$ respectively, demonstrating the non-commutativity of translations.
    (b) 7 translations, $\gamma_1,\ldots,\gamma_7$, denoted as arrows on a 14-gon, which generates the translation group of the $\{14,7\}$ tiling.
    }
    \label{fig:tiling}
\end{figure}

\noindent{\it Wave basis of high-dimensional representations of hyperbolic lattices}---In this section we consider general principles for constructing wave basis from lattice potentials. 
Let's first briefly review the case of Euclidean lattices, where wave eigenstates are described by Bloch's theorem. Because the translation group $T$ of Euclidean lattices is  abelian  and all elements of $T$ commute with the Hamiltonian $H$ (which has the same periodicity of the lattice), one can choose a set of waves that are eigenstates of $H$ and all elements $t_{R}$ in $T$ 
\begin{equation}\label{EQ:translation1}
    t_{R}\psi(r) \equiv\psi(t_R^{-1}r)=  \psi (r) e^{ik R},
\end{equation}
and these waves can be written as
\begin{equation}\label{EQ:Bloch1}
    \psi(r) = e^{-ikr} u(r),
\end{equation}
where $r$ is the position in space, $k$ is the crystal momentum, and $u(r)$ is a function with the same periodicity as that of the lattice. 
This theorem from Bloch  [Eq.~\eqref{EQ:Bloch1}] has an equivalent description, using  the Wannier basis, a complete orthogonal basis that characterizes localized molecular orbitals of crystalline systems~\cite{Wannier1937}, 
\begin{equation}\label{EQ:Wannier1}
   \psi(r)=\sum_R \phi_R(r) e^{-ikR} =\sum_{t_R\in T}  [t_R\; \phi(r)] e^{-ikR},
\end{equation}
where the Wannier function $\phi_R(r)$ obeys $\phi_{R}(r)=t_{R}\phi(r)=\phi(t_R^{-1}r)$. The sum here is over all lattice vectors, (or equivalently all elements of the lattice translation group).
These three formula Eqs.~\eqref{EQ:translation1}-\eqref{EQ:Wannier1} are equivalent to each other.

Next, we generalize this formulation to hyperbolic lattices in the form of $\{p,q\}$ tilings (i.e., lattices of $p$-sided polygons tiling the hyperbolic plane in a regular pattern such that $q$ polygons meet at each vertex).  The space symmetries of these tilings are described by the Coxeter group $G$, a nonabelian infinite group which is analogous to the space group of Euclidean lattices~\cite{coxeter1934discrete}.  For tilings that satisfy the condition that $q$ has a prime divisor less than or equal to  $p$, a \emph{generalized translation subgroup} $T\subset G$ can be defined, where each element $t\in T$ has a one-to-one correspondence with each polygon in the tiling~\cite{Yuncken2003}, and the lattice dual to the tiling (i.e., $\{q,p\}$) can be defined as a \emph{generalized Bravais lattice}.
This algebraic generalization of translations and Bravais lattices recovers the conventional definition when applied to a regular Euclidean Bravais lattices.  Similar to the Euclidean case, lattices that don't satisfy this criterion (non-Bravais lattices) can be considered as a Bravais lattice with a basis (internal DOFs).
This definition differs slightly from the one used in Ref.~\cite{Boettcher2022}, because our translation group is not limited to hyperbolic translations, and it broadens Bloch's theorem to more generic non-Euclidean lattices (e.g., spherical lattices like 600-cells~\cite{coxeter1973regular}).

This generalized translation group enables a generalization of  Bloch's theorem to higher-dimensional irreps.  To achieve this, we start by drawing analogies with Eqs.~\eqref{EQ:translation1} and~\eqref{EQ:Wannier1}.
Here, although the Hamiltonian $H$  commutes with all elements of $T$, the group $T$ itself is nonabelian.   Thus, some eigenstates of $H$ must lie in some high-dimensional irreps of $T$. That is, if $\psi_1(x)$ is an eigenstate of $H$ with energy $E$, there must be an irrep $\rho$ (say $d$ dimensional) and $d-1$ other eigenstates $\psi_2(x),\ldots,\psi_d(x)$ with the same energy $E$ such that $\forall t\in T$
\begin{equation}\label{EQ:translation}
    t \psi_j(r) \equiv \psi_j(t^{-1} r) = \psi_i(r) \rho(t)_{ij},
\end{equation}
where the $d\times d$ matrix $\rho(t)$ is a $d$-dimensional irrep of the translation group $T$~\cite{scott1996linear}.
This is the non-abelian generalization of Eq.~\eqref{EQ:translation1}, and the $d$ degenerate eigenstates $\psi_i(r)$ are the \emph{generalized Bloch waves}. They can be constructed Wannier basis in analogy to Eq.~\eqref{EQ:Wannier1}
\begin{equation}\label{EQ:Wannier}
   \psi_j(r)=\sum_{t\in T} [t\; \phi_i(r)]\rho(t^{-1})_{ij},
\end{equation}
where a set of $d$  Wannier functions $\phi_i(r)$ can be obtained by solving the eigenstates of the Hamiltonian (see the next section). 
It is straightforward to verify that waves constructed via Eq.~\eqref{EQ:Wannier} indeed transform as Eq.~\eqref{EQ:translation} (see SM). 

Below, we show that for each irrep,
all of its eigenmodes can be obtained using this generalized Bloch's theorem. By exploring all  irreps of $T$, a complete description of all eigenmodes can in principle be obtained.

\noindent{\it Eigenmodes in non-Euclidean lattices}---In this section we apply the generalized Bloch theorem to find eigenmodes of any high-dimensional irrep $\rho(t)$.
In general, we can write any states as 
\begin{equation}
|\psi\rangle = \sum_{t\in T}\sum_{a=1}^{n} c(t,a) |v^{(a)}_{t}\rangle .
\end{equation}
Here we label each unit cell using  elements of the translation group $t\in T$ and $|v^{(a)}_{t}\rangle$ is a complete orthogonal basis for the $n$ DOFs (labeled by $a$) in  unit cell $t$.
In the continuum, the index $a$ is a continuous variable labeling  coordinate $r$ (plus some additional indices for internal DOFs).
One convenient choice of basis is to require  $|v^{(a)}_{t}\rangle = t |v^{(a)}_{I}\rangle$, where the basis in the unit cell $I$ at the origin can be chosen arbitrarily as $|v^{(a)}_{I}\rangle=|v^{(a)}\rangle$  and then the basis of any other unit cell is obtained via a  translation. Due to the one-to-one correspondence between group elements of $T$ and unit cells, this approach defines a unique set of basis $|v^{(a)}_{t}\rangle$. 
In this basis, $|v^{(a)}_{t}\rangle$ can effectively be decomposed into the direct product of $|v^{(a)}_{t}\rangle =|v^{(a)}\rangle\otimes |t\rangle$, where $ |t\rangle$ labels the unit cell and $|v^{(a)}\rangle$ spans the linear space of DOFs in a unit cell.
As a result, any Hamiltonian (or dynamical matrix) that preserves the lattice translational symmetry can be written in the following form
\begin{align}
  H=& \sum_{t'\in T} \mathcal{H}_{t'} \otimes \sum_{t\in T}\ket{t}\bra{t\, t'},
\end{align}
where $\mathcal{H}_{t'}$ is an $n\times n$ matrix defined in the linear space of $|v^{(a)}\rangle$.
It describes the hybridization between unit cells $t$ and $t\, t'$. If $H$ is hermitian,  $\mathcal{H}_{t'}=\mathcal{H}^\dagger_{(t')^{-1}}$.

Following the generalized Bloch's theorem discussed above [Eq.~\eqref{EQ:Wannier}], we write the Bloch-wave eigenstates of a $d$-dimensional irrep,
\begin{equation}\label{EQ:latticewave}
    |\psi_j\rangle = \sum_{t\in T}
    \left\lbrack
    \sum_{a=1}^{n}  \lambda_{a,i}\, 
    |v^{(a)}\rangle \otimes |t\rangle 
    \right\rbrack
    \rho(t^{-1})_{ij} .
\end{equation}
Using this construction, the eigenvalue problem $ H |\psi_j\rangle = E |\psi_j\rangle$ is converted to 
the eigenvalue problem of a $d n\times d n$ matrix
$H(\rho)  \lambda = E(\rho) \lambda$ where
\begin{align}
   H(\rho)= \sum_{t\in T} \mathcal{H}_{t} \otimes \rho(t^{-1})^T.
\end{align}
Each eigenvalue give us an eigen-energy, $E$, and the corresponding eigenvector, $\lambda_{a,i}$, yields $d$ degenerate Bloch waves, carrying this $d$-dimensional irrep [Eq.~\eqref{EQ:latticewave}]. It is important to highlight that for each $d$-dimensional irrep, we shall obtain $d\times n$ eigen-energies, i.e., $d \times n$ energy bands, each $d$-fold degenerate ($d^2 \times n$ eigenstates in total).
This is in sharp contrast to  Euclidean lattices, where the band number is determined solely by $n$, because $d=1$.
The fact that $d^2 \times n$ eigenstates emerge here, instead of $n$, is in analogy to the regular representation of a finite group~\cite{scott1996linear}, where a $d$-dimensional irrep reoccurs $d$ times and thus $\sum_\textrm{irreps} d^2=$ the number of group elements.
A similar procedure can be done solving eigenstates in the continuum, as described in the SM.

\begin{figure}
\centering
    \includegraphics[width=0.48 \textwidth]{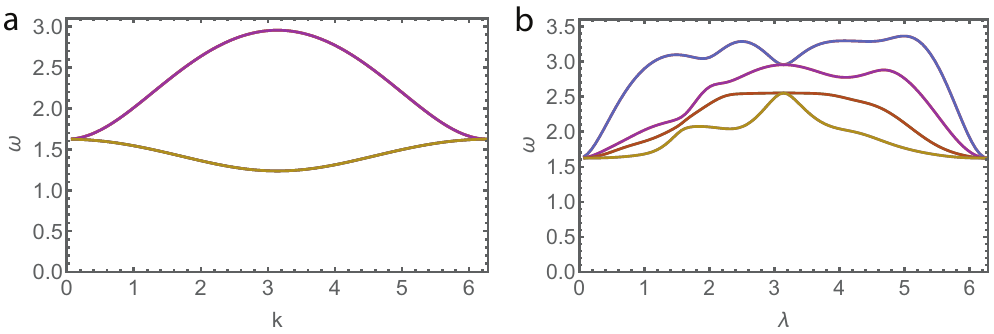}
    \caption{Phonon band structure of a hyperbolic spring network on $\{7,14\}$. 
    (a) A 1D cut of phonon bands from 1D representations, where each wave vector $k$ labels a 1D representation of $T$. Same as in Euclidean lattices, the number of bands ($2$) coincides with the number of DOF per unit cell $n=2$. (b) A 1D cut of phonon bands from 2D irreps, where each $\lambda$ marks one 2D irrep [Eq.~\eqref{EQ:Representation}]. Here, the band number is $d\times n=4$ instead of $n=2$, and each band is 2-fold degenerate. 
     }
    \label{fig:phonon_bands}
\end{figure}

\noindent{\it Phonons on $\{14,7\}$}---We now demonstrate the principles discussed above in a particular hyperbolic lattice: the $\{14,7\}$ tiling (Fig.~\ref{fig:tiling}).
The translation  group $T$ of $\{14,7\}$ can be generated by the hyperbolic translations that translate the central 14-gon to its neighbors, $\{\gamma_i\}_{i=1}^{7}$, as shown in Fig.~1 on the Poincar\'e disk model (which maps the infinite hyperbolic plane to the unit disk $\mathbb{D}$)~\cite{iversen_1992}.  
It is straightforward to see that this is a nonabelian group, i.e., operations $\gamma_i$ do not commute with one another (example shown in Fig.~\ref{fig:tiling}). 
Acting products of $\gamma_i$ generates all 14-gons  without overlapping on the $\{14,7\}$, and they must satisfy two constraints, $\gamma_5\gamma_2\gamma_6\gamma_3\gamma_7\gamma_1\gamma_4=1$ and $\gamma_5\gamma_3\gamma_1\gamma_6\gamma_4\gamma_2\gamma_7=1$. 
By identifying edges $i$ with $i'$, a $14$-gon becomes a genus-3 torus $\Sigma_3$~\cite{levy2001eightfold} and each $\gamma_i$ becomes a loop on this torus, i.e., one element of the fundamental group $\pi_1(\Sigma_3)=\{\langle a_1,a_2,a_3,b_1,b_2,b_3\rangle,[a_1,b_1][a_2,b_2][a_3,b_3]=1\}$ (here $[t,t']\equiv t t' t^{-1} t'^{-1}$ is the commutator between two group elements). Based on this mapping, an isomorphism between $T$ and $\pi_1(\Sigma_3)$ can be obtained (see SM),
utilizing the relation between the deck group of universal covers and fundamental groups~\cite{hatcher2005algebraic}.
Thus we can use $d$-dimensional irreps of the $a$'s and $b$'s to construct irreps of $T$. 

Here, we use an explicit model mechanical system to demonstrate the principles discussed above. We place a mass $m=1$ at each node of $\{7,14\}$ [red dots of Fig.~\ref{fig:tiling}(a)] and use an elastic spring (with spring constant $k=1$) to connect neighboring nodes. This spring network has two (in-plane) degrees of freedom per unit cell ($n=2$), thus for modes in 1D representations of $T$, we  expect two phonon bands, which is indeed what we observe in Fig.~\ref{fig:phonon_bands}(a). Here, 1D representations span a $6$-dimensional BZ (from the 6 generators $a$'s and $b$'s) and we plot a 1D cut of this 6D space using the representation
$a_1=e^{ik}$ and $a_2=a_3=b_1=b_2=b_3=1$.

For higher-dimensional irreps, the BZ is 
generalized to a $[(2g-1)d^2+1]$-dimensional space ($g=3$ for $\{14,7\}$)~\cite{Rapinchuk1996}, which labels all $d$-dimensional irreps. Here, to demonstrate the generalized Bloch's theorem for higher-dimensional irreps, we plot a 1D cut of this 21-dimensional band structure using this set of 2D irreps
\begin{align}\label{EQ:Representation}
    a_{\alpha}=\cos\lambda \; I+i \sin \lambda \; \sigma_{\alpha}, 
    \;
    b_1=b_2=I, 
  \;
    b_3=e^{-i \lambda} I,
\end{align}
where $I$ is the $2\times 2$ identity matrix and $\sigma_{\alpha}$ with ${\alpha}=1,2,3$ are the three Pauli matrices. For $0<\lambda<\pi$ or $\pi<\lambda<2\pi$, each $\lambda$ labels a 2D irrep, and we can compute its eigen-frequencies and eigen-modes following the generalized Bloch theorem [Fig~\ref{fig:phonon_bands}(b), see SM for details]. Indeed, we found $d \times n=4$ phonon bands, each 2-fold degenerate.

\begin{figure}
    \centering
    \includegraphics[width=0.45 \textwidth]{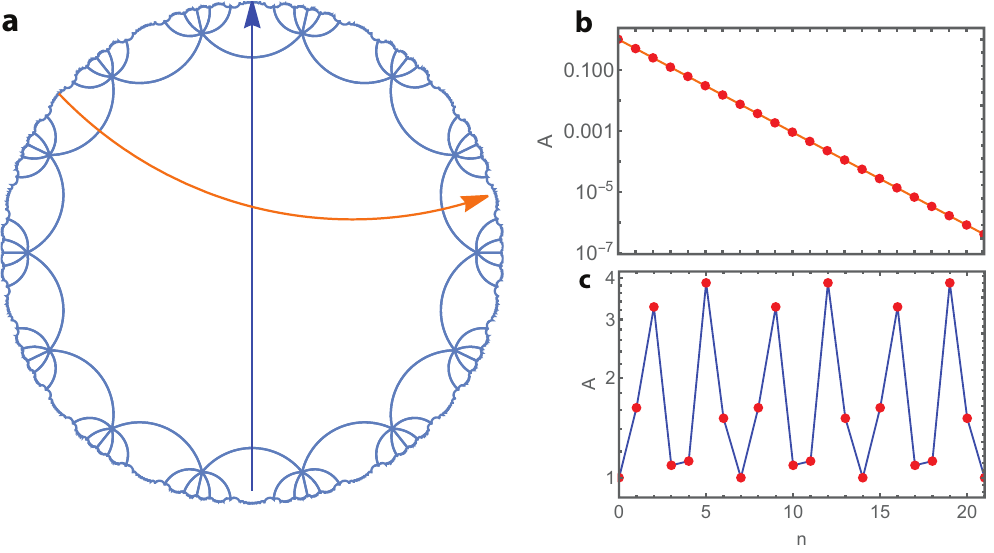}
    \caption{Absence of localized modes in 1D representations. (a) Two types of geodesics in $\{14,7\}$ constructed by repeatedly acting either (i) one generator ($\gamma_7$ in this case, yellow) or (ii) an element in $[T,T]$ (blue). (b,c) Amplitude of a zero mode (see SM) from a non-unitary 1D presentation on discrete degrees of freedom in unit cells for these two geodesics, where  the mode exponentially decays on type (i) geodesics (b), and oscillates on type (ii) geodesics (c). 
    The periodicity in (c) is a universal feature of all 1D representation modes.
    }
    \label{fig:non_unitary}
\end{figure}

\noindent{\it Non-unitary representations and bulk-edge correspondence}---In this section, we discuss states with localized edge modes. Although the origin of such edge modes may vary (e.g., topological or accidental), they all involve non-unitary representations of the translation groups. In Euclidean space, a wave from a non-unitary representation takes the same form as the Bloch wave $\psi(r)=u(r) e^{- i k r}$, but its wavevector $k$ takes a complex value. In general, in an Euclidean lattice, any bulk (edge) modes  can be written as the superposition of unitary (non-unitary) modes, which correspond to a Fourier (Laplace) transformation.

In hyperbolic Bravais lattices, edge modes also involve non-unitary representations, but interestingly, they  cannot be described by 1D non-unitary representations of the translation groups, due to the non-abelian nature of $T$. For any 1D representation, although repeatedly acting one translation $t$ with $\rho(t)^\dagger\ne\rho(t)^{-1}$ does lead to coherent decay (or growth) of its amplitude (Fig.~\ref{fig:non_unitary}b), the mode must be invariant under any translation that belongs to the commutator subgroup of the translation group (Fig.~\ref{fig:non_unitary}c). The commutator subgroup $[T,T]$ is
generated by all commutators $[t,t']$ of group elements of $T$, and for any 1D representations, $\rho([t,t'])=1$, i.e., these modes must be invariant under $[t,t']$ (as shown in the SM).
In contrast to Euclidean lattices, where $[T,T]$ is always trivial, the commutator subgroup of a hyperbolic lattice contains infinitely many translations in various directions, along which 1D-representation modes cannot decay, making it impossible to form a localized edge state. (although certain corner modes is allowed, e.g., at the tip of a sharp wedge along the yellow line in Fig.~\ref{fig:non_unitary}).


This observation has a deep impact on topological edge-bulk correspondence. In Euclidean space, it has long been know that  a nontrivial topological structure in bulk band can lead to nontrivial edge modes, known as the bulk-edge correspondence~\cite{fradkin_2013}. For a hyperbolic lattice, such correspondence necessarily requires higher-dimensional representations.
Even if a bulk topological index only involves 1D representations (e.g.,~Ref.~\cite{Urwyler2022hyperbolic}), the corresponding edge states (if exists) must involve higher dimensional non-unitary irreps, because 1D (unitary and non-unitary) representations cannot form edge modes.
This observation is one example demonstrating the incompleteness of 1D representations in non-Euclidean lattices. In fact, because 1D representations (unitary or not) can only lead to waves invariant under any transition in $[T,T]$, they cannot offer a complete wave basis. Any modes (bulk or edge) not obeying this invariance necessarily require higher-dimensional irreps. 

Another interesting feature that arises by considering non-unitary representations is the existence of zero modes.  Although the mechanical lattice we consider here has coordination number $z=14$, which is far above the Maxwell criterion for stability~\cite{sun2012surface,Lubensky_2015,Mao2018,Sun2020}, the lattice is guaranteed to have zero modes under open boundary conditions.  This can be seen, e.g., in 1D representation, by comparing the number of constraints (7 per unit cell from the springs) and free variables (13 from 6 complex momenta $k_1,...,k_6$ and one for the direction of displacements, see SM for details).  The excess free variables allows zero modes in the linear model, in a way similar to how corner modes arise in over-constrained lattices~\cite{Saremi2018}. This works similarly for 2D irreps where we have 21 complex momenta. An alternative way to see the existence of these zero modes is that the fraction of boundary nodes is $\mathcal{O}(1)$ in hyperbolic lattices, leaving a macroscopic number of removed constraints.

\noindent{\it Conclusions and discussions}---In this paper we generalize Bloch's theorem to high-dimensional irreps of infinite hyperbolic lattices, using linear combinations of Wannier basis.  We find that hyperbolic lattices exhibit a number of unusual features, in contrast to Euclidean lattices, from high degeneracy of band structures to modifications of bulk-edge correspondence that require high-dimensional irreps.  We apply this theory to a model mechanical hyperbolic lattice, and compute its band structure and zero modes.

Same as Bloch's theorem in Euclidean space, once an irrep is given, our theorem enables us to find all
$d^2\times n$ eigenmodes of this irrep.
In parallel to our theorem, another interesting question is to find all irreps and to prove their completeness, which is a nontrivial mathematical problem due to the rich variety of hyperbolic lattices and the nonabelian nature of their translation groups.
In contrast to Euclidean space, where all irreps can be labeled by $k$ points in the BZ, hyberbolic lattices require infinitely many high-dimensional ``BZs" (e.g., for lattices studied in Ref.~\cite{Rapinchuk1996}, $d$-dimensional irreps span a $[(2g-1)d^2+1]$-dimensional space).

A number of interesting new questions arise for future studies. For example, how do acoustic phonon branches show up in this formulation, and what is the new form of Goldstone's theorem~\cite{Nambu1960,goldstone1961field}?  Interestingly, in the mechanical hyperbolic lattice we considered here, 
$k=0$ modes in 1D representation and $\lambda=0$ modes in 2D (reducible) representation all have finite frequencies $\omega>0$, in contrast to acoustic phonon modes in Euclidean lattices, which are protected to have $\omega=0$ at $k=0$ by Goldstone's theorem.  The reason is that uniform translations in hyperbolic lattices are isometries which can be described as boosts in the hyperboloid model, instead of $k=0$ modes.  It would be very interesting to show the new form of Goldstone's theorem.  Furthermore, unusual symmetries and bulk-edge correspondence in hyperbolic lattices also provide a huge space for new topological states.

\noindent{\it Acknowledgements}--- This work was supported in part by the Office of Naval Research (MURI N00014-20-1-2479 N.C., F.S., J.M., K.S. X.M.).

\appendix

\begin{widetext}

\section{Bloch waves in curved space}

\subsection{Translation of generalized Bloch waves}
A $d$-dimensional representation maps each group element to a $d\times d$ matrix, i.e., $t \mapsto \rho(t)$ for $t \in T$, 
and the $d\times d$ matrix $\rho(t)$ obeys 
$\rho(t_1)\rho(t_2)=\rho(t_3)$, if $t_1 t_2= t_3$.
Under the transformation $t$, waves/modes ($\psi_i(r)$) that belong to the representation $\rho$ shall 
transform as 
$t \psi_j(r)=\psi_i(r)\rho(t)_{ij}$, where $i,j=1,2,\ldots,d$.


In the main text we defined the generalized Bloch waves $\phi_i(r)$ of a representation $\rho$ using Wannier functions $\phi_i(r)$. Here, we verify that under translation $t\in T$, these generalized Bloch waves indeed follow this correct group representation $\rho$: 
$t \psi_j(r)=\psi_i(r)\rho(t)_{ij}$.


Under translation $s\in T$, the generalized Bloch wave transforms as
\begin{equation}
 	s\psi_j(r)=\sum_{t\in T} [st\; \phi_i(r)]\rho(t^{-1})_{ij}
 	=\sum_{p\in T} [p\;\phi_i(r)]\rho(p^{-1}s)_{ij}
 	=\sum_{p\in T} [p\;\phi_i(r)]\rho(p^{-1})_{ik}\rho(s)_{kj}
 	=\psi_k(r)\rho(s)_{kj}.\label{SI:EQ:transformationlaw}
\end{equation}
Here we define $p=st$, and the rearrangement theorem ensures that $\sum_p$ sums over all elements of $T$ without repetition. This relation directly verifies that these Bloch waves do belong to the $\rho$ representation of the group $T$.

\subsection{Solving Wannier functions of a given representation}
For tight-binding models (or models with a finite number of DOFs per cell), the generalized Bloch waves can be easily calculated using the matrix (tensor) formula shown in the main text. 

In this section, we consider models defined in continuous space, with infinite DOFs in each unit cell, and we show a systematic approach of solving Wannier functions for a $d$-dimensional irrep $\rho$. 
First, we choose an arbitrary function $\phi_0(r)$, such that $|\phi_0(r)|$ decreases rapidly as $r$ moves away from the unit cell $I$ (the origin). This function allows us to define a $d\times d$ matrix
\begin{equation}
	\Xi(r)=\sum_{t\in T}\ [t\; \phi_0(r)]\rho(t^{-1}). \label{SI:EQ:matrixfunction}
\end{equation}
By choosing proper $\phi_0(r)$, we can make this matrix invertible in a unit cell (where some discrete singular points with $\det \Xi(r)=0$ are allowed).
Using the translation formula shown below, this ensures that $\Xi(r)$ is invertible for any $r$ (up to some unimportant singular points).

Under $t \in T$, this matrix $\Xi(r)$ transforms as
\begin{equation}
	[t\; \Xi(r)]\equiv \Xi(t^{-1}r)=\Xi(r)\rho(t). \label{SI:EQ:matrixtransformation}	
\end{equation}
If $\Xi(r)$ is invertible, we can rewrite the generalized Bloch wave $\psi(r)=(\psi_1(r),...,\psi_d(r))$ as
\begin{equation}\label{SI:EQ:Bloch:formula}
\psi(r) =\psi(r) \Xi^{-1}(r) \Xi(r)= u(r) \Xi(r)
\end{equation}
where $\psi(r)$ is written as a column vector and here we defined a column vector $u(r)=(u_1(r),...,u_d(r))$ as
\begin{align}
u(r)=\psi(r)\Xi^{-1}(r)
\end{align}
The same formula can also be written in terms of their components
\begin{align}
\psi_j(r) = u_i(r) \Xi_{ij}(r) \;\;
\textrm{and} \;\; u_j(r) = \psi_i(r) \Xi^{-1}_{ij}(r)
\end{align}
If $\Xi(r)$ contains singular points, $u(r)$ would diverge at these points, but $\psi\propto u \Xi$ remains non-singular.

This new formula carries the same information, but it has one advantage: $u(r)$ is a periodic function, i.e., it is invariant under any lattice translation $u= t u$. This periodic property can be easily verified using the definition $u$:
\begin{equation}
	t\; u(r)
	=[t\; \psi(r)][t\;\Xi(r)]^{-1}
	=[\psi(r) \rho(t)] [\Xi(r) \rho(t)]^{-1} 
	=\psi (r)\Xi^{-1}(r)=u(r). \label{SI:EQ:periodicvector}
\end{equation}

In terms of the $u$ functions, the eigenequation of generalized Bloch waves $H\psi_j = E \psi_j$ now becomes
\begin{align}
    H[u_i(r)\Xi_{ij}(r)]=Eu_i(r)\Xi_{ij}(r) 
    \label{SI:EQ:eigenequation}
\end{align}
Because $u(r)$ is a periodic function, to solve this eigenvalue problem, we just need to focus on one unit cell using periodic boundary conditions, which dramatically reduces the complexity of the calculation. 

It is worthwhile to emphasize that although the procedure above involves a function $\phi_0$, which we choose arbitrarily, the final results (eigenvalues and Bloch waves) are independent of this choice. More precisely, the functions $u(r)$ do depends on the choice of $\phi_0(r)$, but $u(r)\phi_0(r)$ doesn't depends on it, and the same applies to the Bloch wave $\psi(r)$, which is proportional to $u(r)\phi_0(r)$. 

We conclude this section by revisiting Eq.~\eqref{SI:EQ:Bloch:formula},
\begin{equation}
\psi(r) = u(r) \Xi(r).
\end{equation}
Because $u(r)$ is a periodic function, this formula is the direct generalization of Bloch's theorem. It shows that eigenfunctions can be written as the product of a periodic function $u(r)$ and another function $\Xi(r)$ that carries information  about the representation of the group.

In addition, for an abelian translation group, it is known that for a single energy band, the Wannier function is independent of the representation (i.e., the value of $k$). This conclusion doesn't necessarily generalize to high-dimensional irreps, and thus in general, different irreps may have different Wannier functions.

\section{Phonon modes in a non-Euclidean spring network}

\subsection{The hyperboloid model}
Here we consider the \{14,7\} tiling on a hyperbolic surface. We embed the hyperbolic surface in a 3D Lorentz space with the metric tensor
\begin{align}
g=
	\begin{pmatrix}
		-1 & 0 & 0 \\
		0 & 1 & 0 \\
		0 & 0& 1
	\end{pmatrix},
\end{align}
Here we consider the hyperboloid~\cite{anderson2006hyperbolic} $\mathbb{H}=\{u| \braket{u,u}=-1, u^0>0\}$, where $u$ is the 3D coordinate of the Lorentz space and $\braket{u,v}$ is the inner product. 
On this hyperboloid, the distance between two points $p$ and $q$ is defined as
\begin{equation}
d(p,q)=\arccosh(-\braket{p,q})
\end{equation}
This choice of the hyperboloid and the distance function  $\{\mathbb{H},d\}$ is called the hyperboloid model of a hyperbolic space. Here, we will use this hyperboloid to host our \{14,7\} tilling.

Because this hyperbolic lattice is embedded in a 3D Lorentz space, the space symmetry group here is a subgroup of the $O(1,2)$ Lorentz group. This is why we can write group elements of our lattice translation group $T$ as $3\times 3$ matrices below, e.g., Lorentz boosts and space rotations in a 3D Lorentz space.

\subsection{Translation group of \{14,7\} and explicit isomorphism}
The translations $\gamma_i$ realized in hyperboloid model are
\begin{equation}
\begin{aligned}
	&\gamma_1=R(\frac{2\pi}{7})R(-\frac{\pi}{2})B_x(r_0)R(\frac{\pi}{2})\\
	&\gamma_2=R(\frac{2\pi}{7})R(-\frac{11\pi}{14})B_x(r_0)R(\frac{11\pi}{14})\\
	&\gamma_3=R(\frac{2\pi}{7})R(-\frac{15\pi}{14})B_x(r_0)R(\frac{15\pi}{14})\\
	&\gamma_4=R(\frac{2\pi}{7})R(-\frac{19\pi}{14})B_x(r_0)R(\frac{19\pi}{14})\\
	&\gamma_5=R(\frac{2\pi}{7})R(\frac{5\pi}{14})B_x(r_0)R(-\frac{5\pi}{14})\\
	&\gamma_6=R(\frac{2\pi}{7})R(\frac{\pi}{14})B_x(r_0)R(-\frac{\pi}{14})\\
	&\gamma_7=R(\frac{2\pi}{7})R(-\frac{3\pi}{14})B_x(r_0)R(\frac{3\pi}{14})\\
\end{aligned}	\label{SI:EQ:gammamatrices}	
\end{equation}
where
\begin{equation}
	B_x(r)=
	\begin{pmatrix}
		\cosh(r) & \sinh(r) & 0 \\
		\sinh(r) & \cosh(r) & 0 \\
		0 & 0& 1
	\end{pmatrix}, \,
	R(\theta)=
	\begin{pmatrix}
		1 & 0 & 0 \\
		0 & \cos(\theta) & -\sin(\theta) \\
		0 & \sin(\theta) & \cos(\theta) 
	\end{pmatrix}, \,
	r_0=2\arccosh
	\left(\frac{\cos(\frac{\pi}{7})}{\sin(\frac{\pi}{14})}\right) \label{SI:EQ:definingmatrices}	
\end{equation}
It is easy to realize that $B_x$ here is a Lorentz boost, and $R(\theta)$ is a space rotation.

By identifying edges $i$ with $i'$ of a $14$-gon ($i=1,2,\ldots,7$), the $14$-gon is transformed into a genus-3 torus $\Sigma_3$~\cite{levy2001eightfold}, and each $\gamma_i$ becomes a loop on $\Sigma_3$, which corresponds to one element of the fundamental group of the genus-3 torus $\pi_1(\Sigma_3)=\{\langle a_1,a_2,a_3,b_1,b_2,b_3\rangle,[a_1,b_1][a_2,b_2][a_3,b_3]=1\}$. 
Here, $[t,t']\equiv t t' t^{-1} t'^{-1}$ is the commutator between group elements. For example, the translation $\gamma_1$ is the loop  $b_1^{-1} b_2^{-1} b_1$.
on the genus-3 torus $\Sigma_3$.

This mapping from the translation group $T$ to the fundamental group $\pi_1(\Sigma_3)$ is not yet a isomorphism. Due to the different convention used in the definitions of the translation group and the fundamental group, an extra inverse operation is needed to define the isomorphism, i.e., instead of mapping $\gamma_1$ to $b_1^{-1} b_2^{-1} b_1$, we will map $\gamma_1$ to the inverse of $b_1^{-1} b_2^{-1} b_1$. Utilizing the relation between the deck group of universal covers and fundamental groups~\cite{hatcher2005algebraic}, it is easy to verify that this indeed defines an isomorphism between $\{a_i,b_i\}_{i=1}^3$ and $\{\gamma_i\}_{i=1}^{7}$
\begin{equation}
\begin{aligned}
    &\gamma_1\mapsto(b_1^{-1}b_2^{-1}b_1)^{-1}\\
   &\gamma_2\mapsto(b_1^{-1}b_2b_3^{-1}b_2a_2^{-1}b_2^{-1}b_2^{-1}b_1)^{-1}\\
   &\gamma_3\mapsto(b_1^{-1}b_2b_1a_1^{-1}b_3^{-1}b_1a_1b_1^{-1}a_1^{-1}a_3^{-1}a_2^{-1}b_2^{-1}b_1)^{-1}\\
   &\gamma_4\mapsto(b_1^{-1}b_2b_2a_2b_2^{-1}b_1a_1b_1^{-1}a_1^{-1}a_3^{-1}a_2^{-1}b_2^{-1}b_1)^{-1}\\
   &\gamma_5\mapsto(b_1^{-1}b_2a_2a_3a_1b_1a_1^{-1}b_1^{-1} b_3a_3^{-1}a_2^{-1}b_2^{-1}b_1)^{-1}\\
   &\gamma_6\mapsto(b_1^{-1}b_2a_2a_3a_1b_1a_1^{-1}b_1^{-1}b_3)^{-1}\\
   &\gamma_7\mapsto(b_1^{-1}b_2a_2a_3a_1b_2^{-1}b_1)^{-1}
\end{aligned}\label{SI:EQ:explicitisomorhism}
\end{equation}



\subsection{Phonon modes}
Here we consider the spring-mass network defined in the main text, and use the dynamical matrix $D$ to compute the eigen-frequencies and eigenmodes. 
Here, each unit cell has $2$ degrees of freedom, i.e., the 2D displacement of each node, and each node is connected to $14$ nearest neighbors by springs.

Define $v^{1}=(0,1,0)$, $v^{2}=(0,0,1)$ as the two orthonormal vectors in the tangent space of a node. Same as in Euclidean space, small deformations of a spring network (in the linear response regime) form a linear space with basis $\{\ket{v^{\delta}}\otimes\ket{t}\}$,
and $\Delta u\ket{v^{\delta}}\otimes\ket{t}$ represents the displacement of node $t$ in the direction of $tv^{\delta}$ with amplitude $\Delta u$. At the same time, another linear space can be defined $\ket{e^{n}}\otimes\ket{t}$, which records the spring extensions of all springs, i.e., $\Delta l \ket{e^{n}}\otimes\ket{t}$ represents that the length of the $n$th spring in the unit cell $t$ is increased by $\Delta l$. The geometry and the connectivity of the spring network defines a linear mapping $C$ from node deformation to spring extension
\begin{equation}
C\cdot\Delta u\ket{v^{\delta}}\otimes\ket{t}=-\sum_{n=1}^{7}\alpha_{n}^{\delta}\Delta u\ket{e^{n}}\otimes\ket{t}-\sum_{n=1}^{7}\overline\alpha_{n}^{\delta}\Delta u\ket{e^{n}}\otimes\ket{t\gamma_{n}^{-1}} , \label{SI:EQ:bondextension}
\end{equation}
where $C$ is a matrix of dimension (number of springs)$\times$($2\times$ number of nodes), known as the compatibility matrix, and  $\alpha_{n}^{\delta}=\frac{\braket{v^{\delta},\gamma_{n}x}}{\sinh r_0}$, 
$\overline\alpha_{n}^{\delta}=\frac{\braket{v^{\delta},\gamma_{n}^{-1}x}}{\sinh r_0}$ are the geometric parameters of projections of node displacements to spring extensions.
Written in a compact form, 
\begin{equation}
	C=-\sum_{t\in T}C_0\otimes\ket{t}\bra{t}
	-\sum_{t\in T}\sum_{n=1}^{7}\{C_i\otimes\ket{t}\bra{t \gamma_i}\} ,\label{SI:EQ:Cmat}	
\end{equation}
where 
\begin{equation}
	C_0=\sum_{n=1}^{7}\sum_{\lambda=1}^{2}\alpha_{n}^{\lambda}\ket{e^{n}}\bra{v^{\lambda}}\;\;\textrm{and}\;\;  C_i=\sum_{\lambda=1}^{2}\overline\alpha_{i}^{\lambda}\ket{e^{i}}\bra{v^{\lambda}} . \label{SI:EQ:S11}	
\end{equation}

With the $C$ matrix obtained, the dynamic matrix $D= C^\dagger C$ can be written as 
\begin{equation}
	D=D_0\otimes\sum_{t\in T}\ket{t}\bra{t}+\sum_{j=1}^7D^R_j\otimes\sum_{t\in T}\ket{t}\bra{t\gamma_j}+\sum_{j=1}^7D^L_j\otimes\sum_{t\in T}\ket{t\gamma_j}\bra{t}, \label{SI:EQ:Dmatirx}
\end{equation}
where 
\begin{equation}
\begin{aligned}
	D_0=&-\sum_{n=1}^7\sum_{\delta=1}^2\sum_{\lambda=1}^2(\overline\alpha^{\delta}_n\overline\alpha^{\lambda}_n+\alpha^{\delta}_n\alpha^{\lambda}_n)\ket{v^{\delta}}\bra{v^{\lambda}}
	\\
	D^R_j=&-\sum_{\delta=1}^2\sum_{\lambda=1}^2\alpha_j^{\delta}\overline\alpha_j^{
    \lambda}\ket{v^{\delta}}\bra{v^{\lambda}}
    \\
    D^L_j=&-\sum_{\delta=1}^2\sum_{\lambda=1}^2\overline\alpha_{j}^{\delta}\alpha_{j}^{\lambda}\ket{v^{\delta}}\bra{v^{\lambda}} 
\end{aligned}
    \label{SI:EQ:S13}	
\end{equation}

For Bloch waves defined in the main text, 
\begin{equation}\label{SI:EQ:latticewave}
    |\psi_j\rangle = \sum_{t\in T}
    \left\lbrack
    \sum_{a=1}^{n}  \lambda_{a,i}\, 
    |v^{(a)}\rangle \otimes |t\rangle 
    \right\rbrack
    \rho(t^{-1})_{ij} .
\end{equation}
the eigenvalue problem $D |\psi_j\rangle = \omega^2 |\psi_j\rangle$ becomes
\begin{align}
    D(\rho)_{ab,ij} \lambda_{b,j}=\omega^2 \lambda_{a,i} ,
\end{align}
where
\begin{align}
   D(\rho)= D_0 \otimes I_{d} 
    +\sum_{j=1}^7 D_j^R\otimes \rho(\gamma_j^{-1})^T
    +\sum_{j=1}^7 D_j^L\otimes \rho(\gamma_j)^T ,
\end{align}
is a $n d\times n d$ matrix. Thus, we will get $d\times n$ eigen-frequencies, i.e., $d\times n$ bands. For each eigen-frequency, the corresponding eigenvector $\lambda_{a,i}$ gives us $d$ degenerate Bloch waves [Eq.~\eqref{SI:EQ:latticewave}], and thus in total, we get $d^2 \times n$ eigen-modes.
In Fig.~\ref{fig:phonon_modes}, we presented the deformation fields of two degenerate eigenmodes for the lowest band of the 2D irreducible representation at $\lambda=2$ as an example.

\begin{figure}[h!]
\centering
    \includegraphics[width=0.23 \textwidth]{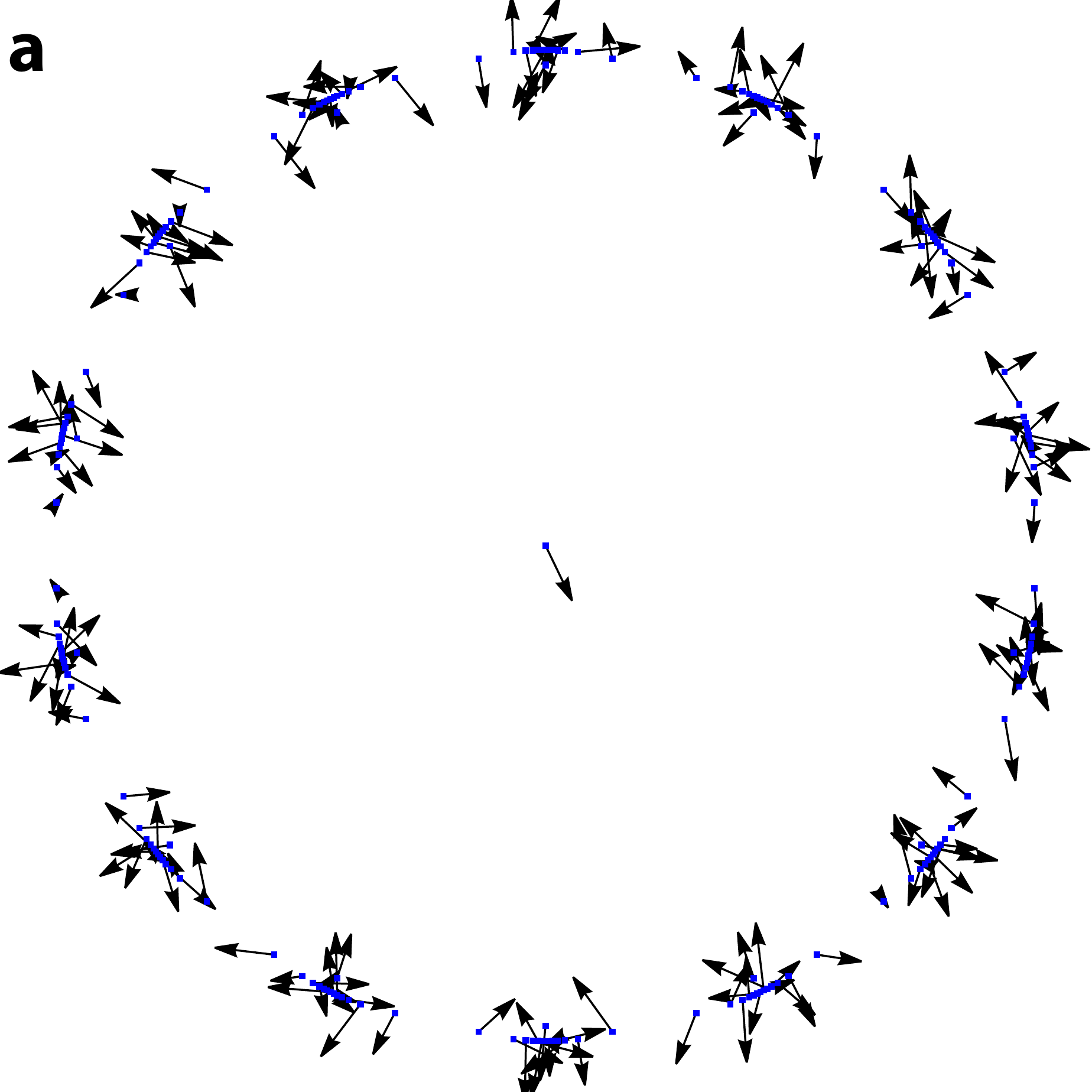}
    \includegraphics[width=0.23 \textwidth]{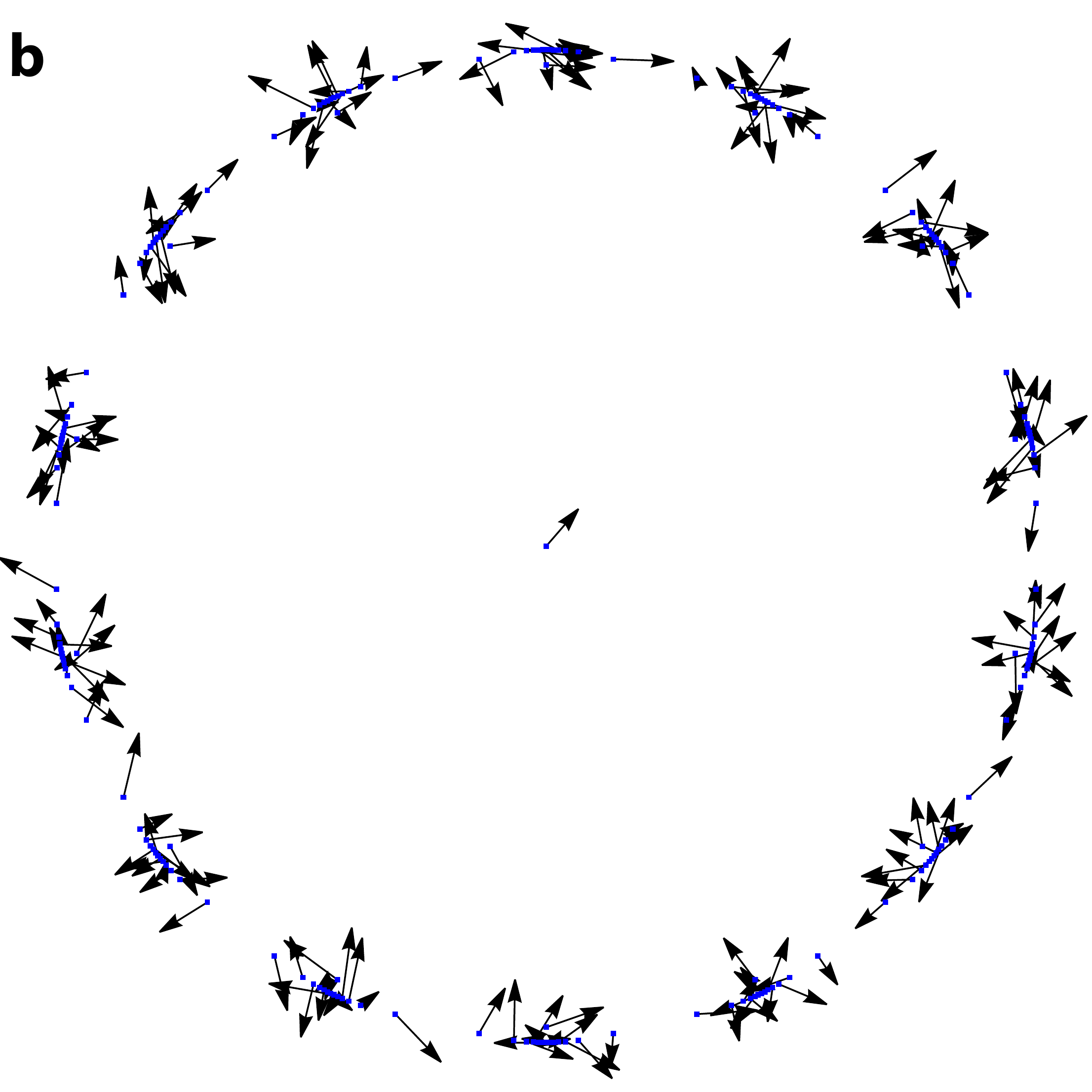}
    \caption{(a,b) Two degenerate eigenmodes at $\lambda=2$ carrying a 2D irrep,  where arrows show displacement vectors of the nodes.
     }
    \label{fig:phonon_modes}
\end{figure}

\subsection{Zero mode in 1D representation}
For a 1D representation $\rho$, $C(\rho)$ is a $7\times 2$ matrix with 6 variables $k_1,...,k_6$, where $\rho(a_j)=e^{i k_j}$, $\rho(b_j)=e^{i k_{j+3}}$. We numerically find that when
\begin{equation}
\begin{aligned}
	&k_1=2.94378-1.55373i, k_2=0.380362-0.383333i, k_3=0.643539+1.23708i\\
	&k_4=1.59631+0.21074i, k_5=0.226622+0.485571i, k_6=-0.169951-0.029206i	
\end{aligned} \label{SI:EQ:solution}
\end{equation}
$C(\rho)$ has a non trivial null space, which is a zero mode.
The amplitude shown in FIG.~3 in the main text is the amplitude of this zero mode.

\subsection{Commutator subgroup}
Consider the commutator subgroup $[T,T]$ of $T$ generated by elements $\{[t,s],s,t\in T\}$ where $[t,s]=tst^{-1}s^{-1}$. 
For any abelian group (e.g., translations in Euclidean space), the commutator subgroup is trivial (i.e., it only contains the identity operator).

In contrast, for hyperbolic lattices, because  $T$ is nonabelian, $[T,T]$ is typically an infinite group with infinitely many group elements (translations). 

One key property of the commutator subgroup lies in the fact that in any 1D representation (unitary or non-unitary), all elements of this subgroup must have a trivial representation $\rho([t,s])=1$. This conclusion can be easily verified
\begin{equation}
	\rho([t,s])=\rho(t)\rho(s)\rho(t^{-1})\rho(s^{-1})=\rho(t)\rho(t^{-1})\rho(s)\rho(s^{-1})=1, \label{SI:EQ:S19}
\end{equation}
because $\rho(s) \rho(t)=\rho(t) \rho(s)$ for any 1D representation. 

This identify implies that  any modes from any 1D representation (unitary or non-unitary) of the translation group must be invariant under any translation $h\in [T,T]$. In other words, the wavefunction of any 1D representation mode must be a periodic function, and any $h\in [T,T]$ (as long as $h\ne I$) is one period of the wavefunction. This observation also means that 1D representations cannot form a complete basis for a hyperbolic lattice, because any modes that don't have such periodicity cannot be written as superposition of 1D representation modes.


One example of such periodicity is shown in Fig.~3 of the main text, where the blue geodesic line follows unit cells $I$, $\gamma_1^{-1}$, $\gamma_1^{-1}\gamma_7^{-1}$,..., $\gamma_1^{-1}\gamma_7^{-1}\gamma_6^{-1}\gamma_5^{-1}\gamma_4^{-1}\gamma_3^{-1}\gamma_2^{-1}, \ldots$. 
Because 
\begin{equation}
	\rho(\gamma_1^{-1}\gamma_7^{-1}\gamma_6^{-1}\gamma_5^{-1}\gamma_4^{-1}\gamma_3^{-1}\gamma_2^{-1})=(\rho(\gamma_1)\rho(\gamma_7)\rho(\gamma_6)\rho(\gamma_5)\rho(\gamma_4)\rho(\gamma_3)\rho(\gamma_2))^{-1}	=(\rho(\gamma_5\gamma_2\gamma_6\gamma_3\gamma_7\gamma_1\gamma_4))^{-1}=1 \label{SI:EQ:S20}
\end{equation}
where the last equation follows from $\gamma_5\gamma_2\gamma_6\gamma_3\gamma_7\gamma_1\gamma_4=1$, we see here that after 7 steps, the amplitude must repeat itself.
Although Fig.~3 is the plot of  a specific mode (the zero mode), this periodicity is a generic feature of any 1D representation modes.

It is worthwhile to note that because geodesics of hyperbolic space diverge, for non-unitary 1D representations, one can still find some direction where the mode decays exponentially. For example, along the yellow geodesic in Fig.~3 generated by repeatedly applying $\gamma_7$.

\end{widetext}




%

\end{document}